\documentclass[reprint,superscriptaddress,amsmath,amssymb,aps,twocolumn,prl,nobibnotes]{revtex4-1}
\usepackage{graphicx}
\usepackage{bm}
\usepackage{amsmath,amssymb}
\usepackage{bbold}
\usepackage[colorlinks=true,citecolor=blue,linkcolor=blue,urlcolor=blue]{hyperref}

\begin{document}
\title{Measuring mixed-state topological invariant in open photonic quantum walk}

\author{Qin-Qin Wang}
\affiliation{Laboratory of Quantum Information, University of Science and Technology of China, Hefei 230026, China}
\affiliation{School of Physics, Hefei University of Technology, Hefei 230009, China}
\affiliation{Anhui Province Key Laboratory of Quantum Network, University of Science and Technology of China, Hefei 230026, China}
\affiliation{CAS Center for Excellence in Quantum Information and Quantum Physics, University of Science and Technology of China, Hefei 230026, China}

\author{Xiao-Ye Xu}
\email{xuxiaoye@ustc.edu.cn}
\author{Yong-Jian Han}
\author{Chuan-Feng Li}
\email{cfli@ustc.edu.cn}
\author{Guang-Can Guo}
\affiliation{Laboratory of Quantum Information, University of Science and Technology of China, Hefei 230026, China}
\affiliation{Anhui Province Key Laboratory of Quantum Network, University of Science and Technology of China, Hefei 230026, China}
\affiliation{CAS Center for Excellence in Quantum Information and Quantum Physics, University of Science and Technology of China, Hefei 230026, China}
\affiliation{Hefei National Laboratory, University of Science and Technology of China, Hefei 230088, China}

\begin{abstract}
Pure-state manifestations of geometric phase are well established and have found applications across essentially all branches of physics, yet their generalization to mixed-state regimes remains largely unexplored experimentally. The Uhlmann geometric phase offers a natural extension of pure-state paradigms and can exhibit a topological character. However, observation of this invariant is impeded by the incompatibility between Uhlmann parallel transport and Hamiltonian dynamics, as well as the difficulty of preparing topologically nontrivial mixed states. To address this challenge, we report an experimentally accessible protocol for directly measuring the mixed-state topological invariant. By engineering controlled nonunitary dynamics in a photonic quantum walk, we prepare topologically nontrivial mixed states from a trivial initial state. Then, by machine learning the full density matrix in momentum space, we directly extract the quantized geometric phase of the nontrivial mixed states. Furthermore, we observe that this transition of the quantized mixed-state topological invariant from a trivial to a nontrivial value is closely associated with the emergence of a topological kink structure. These results highlight a geometric phase framework that naturally extends to open quantum systems both in and out of thermal equilibrium.
\end{abstract}

\maketitle
\raggedbottom

\section*{Introduction}

Geometric phases underpin a wide range of physical phenomena in polarization optics and condensed matter physics \cite{Xiao2010,Cohen2019,Cisowski2022}. In particular, the exemplary Berry geometric phase \cite{Berry1984} has become a cornerstone for identifying the topological properties of quantum systems, including topological insulators and superconductors \cite{Zak1989}. When system-environment coupling is nonnegligible, however, the pure-state geometric phase framework ceases to apply: Open quantum systems are intrinsically mixed, and thermal fluctuations as well as dissipative effects should be incorporated \cite{Markus2012}. This has motivated extensive efforts to generalize geometric phases beyond the pure-state limit \cite{Uhlmann1986,Sjoqvist2000}, especially as the exploration of topological matter moves beyond the idealized zero-temperature regime.

For the topological characterization of open quantum systems, a variety of complementary approaches \cite{Diehl2011,Rivas2013,Viyuela2014,Viyuela2015,Budich2015,Mera2017,Bardyn2018,He2018,Hou2023,Xin2025a} have been developed to extend the Berry geometric phase and topological invariants to mixed states, building on the pioneering work of Uhlmann \cite{Uhlmann1986}. For one-dimensional (1D) topological systems, although measurements of pure-state geometric phases have been demonstrated in quantum simulators \cite{Leek2007,Atala2013,Cardano2017,Jiawei2023}, directly accessing the mixed-state geometric phases \cite{Viyuela2014,Viyuela2015} remains experimentally challenging because the exact Uhlmann parallel-transport process is generally incompatible with the Hamiltonian dynamics \cite{GuoHao2020}. As a result, the time-evolution operators must be carefully engineered to satisfy a relaxed version of the Uhlmann process using auxiliary systems \cite{Hou2021}, and such constructions are known only for several specific models \cite{Viyuela2018,Mastandrea2026}.

Nevertheless, three central challenges remain in advancing the experimental exploration of nontrivial mixed-state topology. First, the preparation of a topologically nontrivial state from an initially trivial state is fundamentally obstructed under coherent unitary dynamics, because of dynamically induced symmetry breaking in the time-evolving state \cite{McGinley2018,McGinley2019,Reid2022}. Second, experimentally determining the quantized Uhlmann geometric phase of the resulting nontrivial mixed state remains demanding. Existing interferometric protocols rely on prior specification of the target density-matrix path and state-specific control within an enlarged system-ancilla space, making them difficult to apply directly to a broader class of mixed states generated in open quantum systems. Third, the physical and geometric mechanisms underlying the transition of the quantized Uhlmann geometric phase \cite{Viyuela2014,Viyuela2015} remain to be resolved experimentally.

Here, we present a direct measurement scheme for the Uhlmann geometric phase viewed as an invariant characterizing the topological properties in an open quantum walk (QW). We first prepare a mixed state with nontrivial topology through slow-quench nonunitary dynamics from a trivial initial state. The direct measurement of the mixed-state geometric phase is achieved by precise control of our experimental platform that grants full tomographic access to the density matrix in momentum space. We observe that the measured mixed-state topological invariant of the dephased final state agrees with the nontrivial topology of the target final Hamiltonian, thereby resolving the usual topological mismatch between the Hamiltonian and the state encountered in unitary dynamics. In addition, we reveal that the emergence of a topological kink structure in the Uhlmann holonomy provides a geometric signature of the critical change in the quantized Uhlmann geometric phase.

\section*{Results}

\subsection*{Mixed-state geometric phase in open QW}
\begin{figure}[t]
\centering
\includegraphics[width=\linewidth]{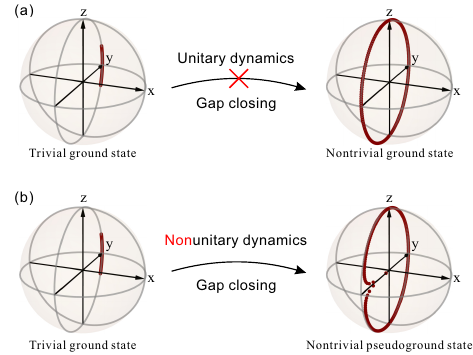}
\caption{\textbf{Protocol for preparing topologically nontrivial states.} \textbf{(a)} Directly preparing a nontrivial ground state with \(\Phi_{\mathrm{B}} = \pi\) from a trivial one with \(\Phi_{\mathrm{B}} = 0\) is not feasible under unitary dynamics. \textbf{(b)} A nontrivial pseudoground state with \(\Phi_{\mathrm{U}} = \pi\) can be achieved from a trivial ground state through almost-adiabatic nonunitary dynamics, except for a small density of excitations near the gap-closing point. Red dots on Bloch sphere denote the momentum-resolved density matrices \(\rho_{k}\) across the BZ, from which the geometric phase is directly extracted.}
\label{fig.Sketch}
\end{figure}

We use a split-step version of QW \cite{Kitagawa2010,Asboth2013,Bo2018,Lin2022} to simulate the Su-Schrieffer-Heeger model of a bipartite lattice featuring two topologically distinct phases. In this protocol, a quantum walker hops on a 1D discrete lattice depending on its internal spin state \cite{Su2019}. The unitary operator for a single time step is given by \(U = R_{y}(\frac{\theta_{1}}{2})S_{\downarrow}R_{y}(\theta_{2})S_{\uparrow}R_{y}(\frac{\theta_{1}}{2})\). Here, \(R_{y}(\theta)\) performs a rotation of the walker's internal spin around the \(y\) axis by an angle of \(\theta\), while the conditional shift operators \(S_{\uparrow} = \sum_{x}\left| x + 1 \right\rangle\left\langle x \right| \otimes \left| \uparrow \right\rangle\left\langle \uparrow \right| + \left| x \right\rangle\left\langle x \right| \otimes \left| \downarrow \right\rangle\left\langle \downarrow \right|\) and \(S_{\downarrow} = \sum_{x}\left| x \right\rangle\left\langle x \right| \otimes \left| \uparrow \right\rangle\left\langle \uparrow \right| + \left| x - 1 \right\rangle\left\langle x \right| \otimes \left| \downarrow \right\rangle\left\langle \downarrow \right|\) move the walker to adjacent sites (labeled by \(\left| x \right\rangle\)) based on its spin state (labeled by \(\{\left| \uparrow \right\rangle,\left| \downarrow \right\rangle\}\)). In momentum \(k\) space, the effective Hamiltonian of QW takes the form (see Materials and Methods for details)

\begin{equation}\label{eq.Hamiltonian}
\begin{aligned}H(\theta_{1},\theta_{2}) &= \int_{- \pi}^{\pi}H_{k}dk \\ &= \int_{- \pi}^{\pi}{\lbrack E_{k}\mathbf{n}_{k} \cdot \overrightarrow{\sigma}\rbrack} \otimes |k\rangle\langle k|dk,\end{aligned}
\end{equation}

where \(\overrightarrow{\sigma} = (\sigma_{x},\sigma_{y},\sigma_{z})\) is the vector of Pauli matrices, \(2E_{k}\) is the energy gap of the Bloch Hamiltonian \(H_{k}\), and \(\mathbf{n}_{k}\) is the unit eigenvector. The effective Hamiltonian respects particle-hole symmetry (PHS), time-reversal symmetry (TRS), and chiral symmetry (CS), with the chiral operator \(\Gamma = \sigma_{x}\) constraining \(\mathbf{n}_{k}\) to lie in the \(y - z\) plane (Fig.~\ref{fig.Sketch}a).

For QW at zero temperature, the topology can be characterized by the quantized Berry geometric phase, obtained by integrating over the first Brillouin zone (BZ)

\begin{equation}\label{eq.Berry}
\Phi_{\mathrm{B}} = i\int_{- \pi}^{\pi}\langle n_{k}|\partial_{k}|n_{k}\rangle dk.
\end{equation}

Accordingly, the QW exhibits a topologically nontrivial phase with \(\Phi_{\mathrm{B}} = \pi\) and a trivial phase with \(\Phi_{\mathrm{B}} = 0\), depending on the control parameters \(\{\theta_{1},\theta_{2}\}\) of the effective Hamiltonian. For an open QW coupled to the environment, the state is generally described by a density matrix rather than a wave function. Its topology can be identified by the Uhlmann geometric phase constructed for the mixed state \cite{Viyuela2014,Viyuela2015}. For the \(2 \times 2\) density matrix \(\rho_{k}\) in momentum space, the Uhlmann connection \(A_{\mathrm{U}} = \lbrack\partial_{k}\rho_{k},\rho_{k}\rbrack\). With this connection, the Uhlmann holonomy over the first BZ takes the form of \(V = \mathcal{P}\mathrm{e}^{\int_{- \pi}^{\pi}A_{\mathrm{U}}}\), where \(\mathcal{P}\) stands for the path ordering operator. The associated Uhlmann geometric phase then reads as

\begin{equation}\label{eq.Uhlmann}
\Phi_{\mathrm{U}} = \arg \lbrack\operatorname{Tr}(\rho_{k_{0}}\mathcal{P}\mathrm{e}^{\int_{- \pi}^{\pi}A_{\mathrm{U}}})\rbrack.
\end{equation}

where \(\rho_{k_{0}}\) represents the mixed state at the initial point of the first BZ with \(k_{0} = - \pi\). \(\arg \lbrack \cdot \rbrack\) and \(\operatorname{Tr}( \cdot )\) denote the argument and the trace, respectively. Thus, given \(\rho_{k}\) across the first BZ, the quantized Uhlmann geometric phase can be directly extracted using Eq.~\eqref{eq.Uhlmann}.

\subsection*{Preparing topologically nontrivial mixed states}
The ground state of the flat-band trivial Hamiltonian with \(\Phi_{\mathrm{B}} = 0\) is easily prepared, and the one of any topologically trivial Hamiltonian can be reached through adiabatic unitary evolution without inducing energy-band transitions \cite{Albash2018}. However, preparing a topologically nontrivial state from a trivial one via unitary dynamics is fundamentally obstructed \cite{McGinley2018,McGinley2019,Reid2022,Barbarino2020}, as illustrated in Fig.~\ref{fig.Sketch}a. If the time-dependent Hamiltonian \(H(t)\) is ramped from a trivial to a nontrivial regime, then it must pass through a gap-closing phase transition. At this critical point, the evolution necessarily becomes nonadiabatic, irrespective of the quench velocity. Although the driven Hamiltonian itself retains all three symmetries, the unitary evolution will break the TRS and chiral symmetry of the time-evolving state \cite{McGinley2018}. Consequently, the evolved state retains only PHS and remains topologically trivial [\(\Phi_{\mathrm{B}}(t) \equiv 0\)], even when the driven Hamiltonian itself becomes nontrivial (see Materials and Methods for details).

To prepare a topologically nontrivial state from an easily initialized trivial one, we impose a noisy term into the quench dynamics \cite{Ying2016}

\begin{equation}\label{eq.dephasing}
\partial_{t}\rho_{k} = - i\lbrack H_{k},\rho_{k}\rbrack + \gamma_{k}\lbrack{\widetilde{\sigma}}_{k}^{z}\rho_{k}{\widetilde{\sigma}}_{k}^{z} - \rho_{k}\rbrack,
\end{equation}

where \({\widetilde{\sigma}}_{k}^{z}(t) = \mathbf{n}_{k}(t) \cdot \overrightarrow{\sigma}\) is the Pauli-\(z\) operator in the instantaneous eigenbasis of the Bloch Hamiltonian \(H_{k}(t)\). In this basis, \({\widetilde{\sigma}}_{k}^{z}(t)\) is diagonal and damps only the off-diagonal elements of the density matrix. Thus, the noisy term induces pure dephasing between the ground and excited states at a rate \(\gamma_{k}(t)\) while preserving the energy band occupation. For the final dephased state, the density matrix is diagonal in the eigenbasis of the final Hamiltonian \(H_{f}\)

\begin{equation}\label{eq.rhode}
\rho_{k}^{\text{de}} = \frac{1}{2}\lbrack\mathbb{1} + {\widetilde{n}}_{k}^{z}{\widetilde{\sigma}}_{k}^{z}(t_{f})\rbrack,
\end{equation}

and has purity \({|{\widetilde{n}}_{k}^{z}|}^{2}\). Here, \({\widetilde{n}}_{k}^{z} = 2P_{g \rightarrow e}^{k} - 1\) denotes the population imbalance between the excited and ground states, where \(P_{g \rightarrow e}^{k}\) is the excitation probability during the quench dynamics (see Materials and Methods for details). For a small quench velocity, the excitation is appreciable only near the energy-gap--closing points and is exponentially suppressed away from them. Consequently, \({\widetilde{n}}_{k}^{z} \approx - 1\) throughout the BZ except in a narrow region around the energy-gap--closing points, where it can approach +1.

Specifically, because the eigenstates of the time-dependent Hamiltonian \(H(t)\) respect all three symmetries, so does the dephased state \(\rho_{k}^{\text{de}}\), as a classical mixture of these eigenstates, inherits the three symmetries. Thus, while the TRS and chiral symmetries are generally broken under unitary dynamics \cite{McGinley2018,McGinley2019,Reid2022}, we find that these dynamically broken symmetries can be restored in the dephased state \(\rho_{k}^{\text{de}}\) under nonunitary dynamics. Crucially, in the limit of almost-adiabatic quench dynamics with a small quench velocity \(\upsilon\), the quantized Uhlmann geometric phase of the dephased final state corresponds to the Berry phase of the final Hamiltonian (see Materials and Methods for details)

\begin{equation}\label{eq.phase-limit}
\begin{gathered}\Phi_{\mathrm{U}}^{\text{de}} = \arg \lbrack\cos (|\overrightarrow{U}|)\rbrack,\\ |\overrightarrow{U}| \approx |\Phi_{\mathrm{B}}^{H_{f}} + \mathcal{O}(\upsilon^{\frac{1}{2}})|.\end{gathered}
\end{equation}

This can be regarded as the nonunitary preparation of a pseudoground state of the nontrivial Hamiltonian \(H_{f}\) from a trivial ground state of \(H_{i}\), with only a small number of excitations due to the slow, almost-adiabatic character of the quench, as illustrated in Fig.~\ref{fig.Sketch}b.

\begin{figure}[t]
\centering
\includegraphics[width=\linewidth]{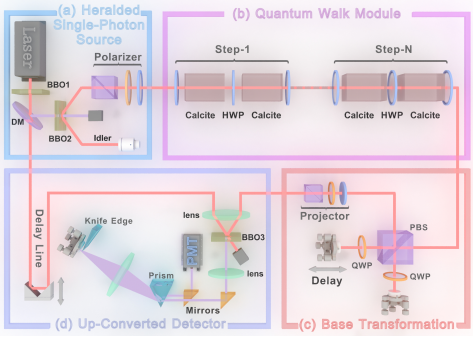}
\caption{\textbf{Experimental QW setup.} The apparatus comprises four functional modules: \textbf{(a)} Photon pairs are produced via spontaneous parametric downconversion in BBO2, where the idler photon serves as a herald for the signal photon entering the QW module. \textbf{(b)} QW is implemented using a sequence of polarization rotations and polarization-dependent time shifts, implemented with HWPs and calcite beam displacers, respectively. In each step, HWPs are mounted on motorized rotation stages that introduce slow, controlled angle fluctuations to emulate dephasing noise. \textbf{(c)} A Michelson interferometer performs the basis transformation and enables interference measurements between different time bins (sites). \textbf{(d)} The walker's position distribution is lastly obtained via single-photon frequency upconversion, allowing for site-resolved detection. BBO, \(\beta - \text{BaB}_{2}\mathcal{O}_{4}\); DM, dichroic mirror; PBS, polarizing beam splitter; HWP, half-wave plate; QWP, quarter-wave plate; PMT, photomultiplier tube.}
\label{fig.setup}
\end{figure}

\begin{figure*}[t]
\centering
\includegraphics[width=0.92\textwidth]{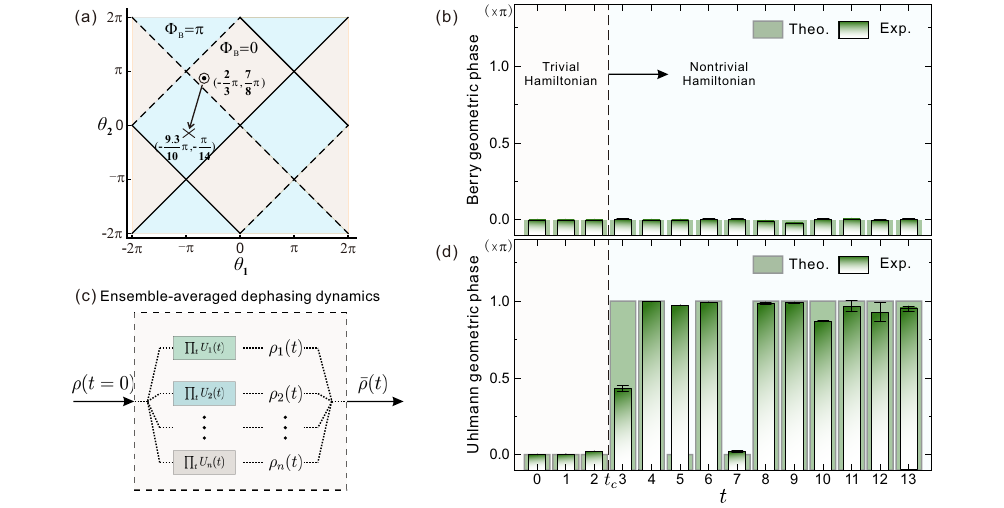}
\caption{\textbf{Experimental strategy and results.} \textbf{(a)} Quench trajectory in the phase diagram, where the system is driven from a topologically trivial Hamiltonian (peach-pink region) to a nontrivial one (blue region). The topologically distinct phases are separated by boundaries at which the energy gap closes at \(E = 0\) (black dashed lines) or \(E = \pi\) (black solid lines). \textbf{(b)} Measured Berry geometric phase (green opaque bars) of the time-evolving pure state under coherent quench dynamics. \textbf{(c)} Realization of dephasing noise through ensemble-averaged dynamics over disordered Hamiltonians. \textbf{(d)} Measured Uhlmann geometric phase (green opaque bars) of the time-evolving mixed state under noisy quench dynamics. In (b) and (d), green transparent bars indicate theoretical predictions. Error bars are obtained from Monte Carlo simulations incorporating photon-counting statistics.}
\label{fig.together}
\end{figure*}

\subsection*{Experimental realization of open QW}
We implement the photonic QW using the setup illustrated in Fig.~\ref{fig.setup} based on the time-bin encoding of the walker's positions and the polarization encoding of the walker's internal spin \cite{Manouchehri2014}. Heralded single photons serve as walkers (Fig.~\ref{fig.setup}a), with the spin basis \(\{\left| \uparrow \right\rangle,\left| \downarrow \right\rangle\}\) mapped to the horizontal and vertical polarizations \(\{\left| H \right\rangle,\left| V \right\rangle\}\) and lattice sites \(\left| x \right\rangle\) mapped to discrete arrival times of photons. Each step of the time-dependent QW evolution is governed by the unitary operator \(U(t) = R_{y}(\frac{\theta_{1}(t)}{2})S_{\downarrow}R_{y}(\theta_{2}(t))S_{\uparrow}R_{y}(\frac{\theta_{1}(t)}{2})\), as shown in Fig.~\ref{fig.setup}b. This operator is physically realized using three half-wave plates (HWPs) for spin rotation \(R_{y}(\theta_{1,2}(t))\) and two calcite crystals for the conditional shift operator \(S_{\uparrow , \downarrow}\). Each calcite crystal introduces a temporal walk-off of \(\tau \simeq 5\) ps between \(\left| H \right\rangle\) and \(\left| V \right\rangle\). After \(\mathcal{N}\) steps, the walker occupies \(2\mathcal{N} + 1\) time bins separated by the interval \(\tau\), and the number of steps \(\mathcal{N}\) sets the effective quench velocity \(\upsilon = 1/\mathcal{N}\).

Next, we experimentally implement the dephasing noise mechanism in Eq.~\eqref{eq.dephasing} by the ensemble-averaged dynamics with spectral disorder \cite{Kropf2016}. In each realization, the rotation parameter \(\theta_{1}\) fluctuates slowly \cite{Schreiber2011,Tang2022} and is sampled from the interval \(\lbrack\theta_{1} - \delta\theta_{1},\theta_{1} + \delta\theta_{1}\rbrack\). Averaging over many realizations yields the desired nonunitary dynamics and mixed state. In momentum space, this procedure produces effective single-qubit dephasing of \(\rho_{k}(t)\): Each disordered Hamiltonian corresponds to a unitary rotation about the \(z\) axis in the instantaneous Hamiltonian's eigenbasis, and averaging over these random rotations suppresses the coherence (see Supplementary Materials). Here, the fluctuation amplitude \(\delta\theta_{1}\) controls the spread of the spectral-disorder distribution and hence the variation of the rotation angles among different disordered realizations. A broader distribution of random rotation angles could lead to stronger suppression of the interband coherence after ensemble averaging. In this way, \(\delta\theta_{1}\) determines the effective dephasing strength described by the dephasing rate \(\gamma_{k}\) in Eq.~\eqref{eq.dephasing} (see Materials and Methods for details).

\begin{figure*}[t]
\centering
\includegraphics[width=0.9\textwidth]{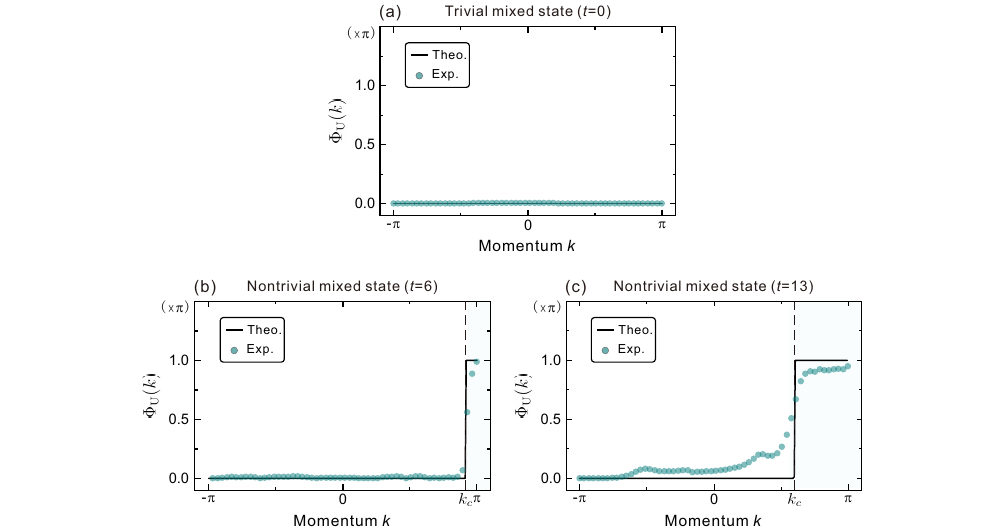}
\caption{\textbf{Experimental topological phase transitions.} Measured Uhlmann geometric phase \(\Phi_{\mathrm{U}}^{k}\) (green circles) versus momentum \(k\) for \textbf{(a)} a topologically trivial mixed state at \(t = 0\) and \textbf{(b and c)} topologically nontrivial mixed states at \(t = 6\) and \(t = 13\) under noisy quench dynamics. The momentum space \(k \in \lbrack - \pi,\pi\rbrack\) is discretized into 58 equally spaced points. The black solid lines denote the corresponding theoretical predictions.}
\label{fig.kink}
\end{figure*}

Figure~\ref{fig.setup}c shows the Michelson interferometer with a tunable arm-length difference, which enables full quantum-state tomography \cite{Wang2024} and thus the extraction of the geometric phase of the time-evolving state. The interferometer is composed of a polarizing beam splitter (PBS), two mirrors, and two quarter-wave plates (QWPs) at \(45^{\circ}\). One mirror is mounted on a piezoelectric transducer for active phase stabilization \cite{Corsi2015}, and the other on a translation stage to precisely control the relative path length. By adjusting the movable mirror, we introduce programmable delays so that photons in \(\left| H \right\rangle\) arrive \(\tau \simeq 5\) ps earlier or later than those in \(\left| V \right\rangle\). After the interferometer, a QWP-HWP-PBS polarization projector performs single-qubit Pauli measurements in the \({\widehat{\sigma}}_{x}\), \({\widehat{\sigma}}_{y}\), and \({\widehat{\sigma}}_{z}\) bases. Interference measurements between time bins at sites \(x\) and \(x + i\) (\(i = 0, \pm 1,\ldots, \pm \mathcal{N}\)) yield \(2\mathcal{N} + 1\) sets of photon-count distributions. Operationally, these measurements correspond to projections onto \(\{\left| H \right\rangle \otimes \left| x \right\rangle\},\{\left| V \right\rangle \otimes \left| x \right\rangle\}\), \(\{\frac{1}{\sqrt{2}}(\left| H \right\rangle \otimes \left| x \right\rangle \pm i\left| V \right\rangle \otimes \left| x^{'} \right\rangle)\}\), and \(\{\frac{1}{\sqrt{2}}(\left| H \right\rangle \otimes \left| x \right\rangle \pm \left| V \right\rangle \otimes \left| x^{'} \right\rangle)\}\) for \(x,x^{'} = 0, \pm 1,\cdots, \pm \mathcal{N}\).

The photon-count statistics are obtained using the upconverted detector shown in Fig.~\ref{fig.setup}d. A spatial delay line equipped with a movable prism mirror scans a 300-mW pump beam, enabling selective upconversion of photons in different time bins separated by 5 ps. The upconverted photons are spectrally filtered using a dispersion-prism system to suppress background noise and are lastly detected by a photomultiplier tube (PMT), yielding all \(2\mathcal{N} + 1\) photon-count datasets.

\subsection*{Measuring mixed-state topological invariant}
We begin by experimentally confirming that a topologically nontrivial state cannot be generated from a trivial initial state through unitary evolution \cite{McGinley2018,McGinley2019,Reid2022,Barbarino2020}. To this end, we implement the slow-quench protocol illustrated by the black arrow in Fig.~\ref{fig.together}a. The system is initialized in the ground state of a topologically trivial Hamiltonian \(H_{i}\), which we prepare by adiabatically evolving the localized real-space state \(\left| x = 0 \right\rangle \otimes \frac{1}{\sqrt{2}}(\left| H \right\rangle - i\left| V \right\rangle)\) \cite{Xu2020}. In momentum space, this localized state corresponds to the ground state of the trivial flat-band Hamiltonian with \(\theta_{2} = \pi\) and a momentum-independent energy. We then ramp the control parameters according to \(\theta_{1,2}(t) = \theta_{1,2}^{i} + (\theta_{1,2}^{f} - \theta_{1,2}^{i})\upsilon t\) with \(\upsilon = 1/13\), thus driving the trivial initial Hamiltonian \(H_{i}\) toward a target nontrivial Hamiltonian \(H_{f}\).

To extract the Berry geometric phase \(\Phi_{\mathrm{B}}(t)\) of the time-evolving pure state, we perform discrete-time-resolved wave function tomography \cite{Xu2018}. For the real-space wave function \(\left| \Psi(t) \right\rangle = \sum_{v}\psi_{v}(t)\left| v \right\rangle\) in the basis \(\left| v \right\rangle = \left| xu \right\rangle\,(u = \uparrow , \downarrow )\), we use a global numerical optimization to determine the amplitudes \(\psi_{v}(t) = \sqrt{p_{v}}(t)e^{i\phi_{v}(t)}\) that best reproduce all photon-counting distributions measured across the \(2\mathcal{N} + 1\) projection bases. Fourier transforming \(\psi_{v}(t)\) yields \(\psi_{k}(t)\) for each \(k\) in the first BZ, from which \(\Phi_{\mathrm{B}}(t)\) is evaluated \cite{Fukui2005}. As shown by the green opaque bars in Fig.~\ref{fig.together}b, we find \(\Phi_{\mathrm{B}}(t) \simeq 0\) throughout the unitary dynamics, even when \(H(t)\) is slowly quenched into the topologically nontrivial regime, demonstrating the infeasibility of generating a nontrivial state from a trivial one via unitary dynamics.

\begin{figure*}[t]
\centering
\includegraphics[width=0.92\textwidth]{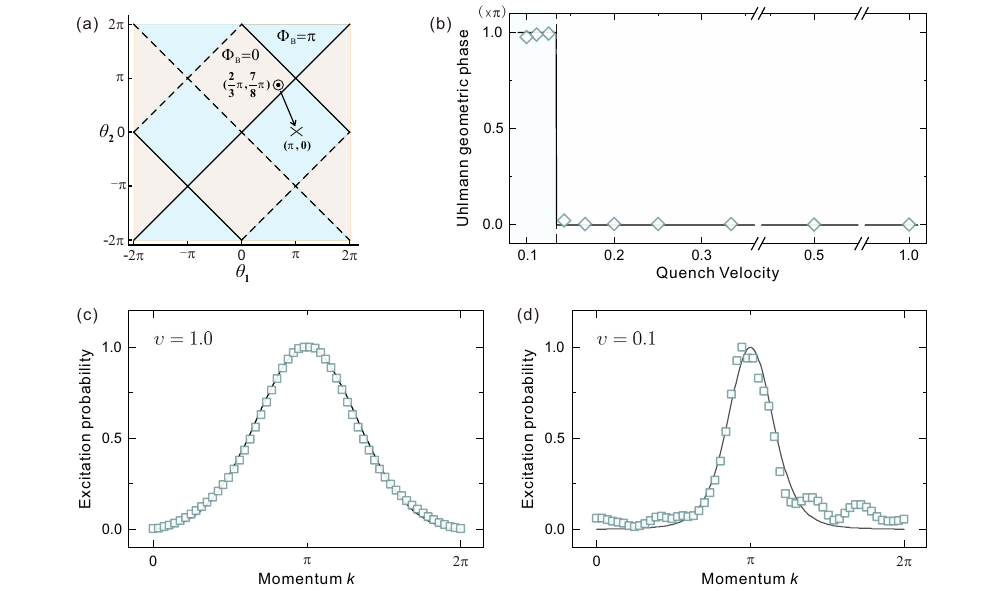}
\caption{\textbf{Experimental characterization of quench-velocity effects.} \textbf{(a)} Quench trajectory in the phase diagram. \textbf{(b)} Measured Uhlmann geometric phase (green diamonds) of the dephased final states as a function of the quench velocity \(\upsilon\). \textbf{(c and d)} Measured excitation probability to the upper band (green squares) of the dephased final states for \(\upsilon = 1.0\) and \(\upsilon = 0.1\), respectively. The momentum space \(k \in \lbrack 0,2\pi\rbrack\) is discretized into 58 equally spaced points. The black solid lines in (b) to (d) denote the corresponding theoretical predictions. Error bars considering the statistical noise are smaller than the symbol size.}
\label{fig.velocity}
\end{figure*}

To realize a topologically nontrivial state, we introduce ensemble-averaged dephasing noise during the quench, as sketched in Fig.~\ref{fig.together}c. Experimentally, this is implemented by randomly varying \(\theta_{1}\) across \(n = 21\) settings within the interval \(\lbrack\theta_{1} - \delta\theta_{1},\theta_{1} + \delta\theta_{1}\rbrack\) with \(\delta\theta_{1} = 0.2\) and performing an ensemble average over all realizations. Under such nonunitary dynamics, reconstructing the full mixed state \(\varrho(t) = \sum_{v,v^{'}}^{}\varrho_{v,v^{'}}(t)\left| v \right\rangle\langle v^{'}|\) at each time step \(t\) is challenging: The \(2\mathcal{N} + 1\) photon-counting collection, although complete for pure-state tomography, is insufficient for directly determining \(\varrho(t)\). To overcome this, we adopt the neural-network ansatz \cite{Wang2024} to parametrize the density matrix \(\varrho_{\boldsymbol{\lambda},\boldsymbol{\mu}}(t) = \sum_{v,v^{'}}^{}\varrho_{v,v^{'}}^{\boldsymbol{\lambda},\boldsymbol{\mu}}(t)\left| v \right\rangle\langle v^{'}|\) with trainable hyperparameters \(\{\boldsymbol{\lambda},\boldsymbol{\mu}\}\). These parameters are optimized by gradient descent to maximize the likelihood of all \(2\mathcal{N} + 1\) photon-counting datasets, yielding the best-fit density matrix \(\varrho_{\boldsymbol{\lambda}^{\ast},\boldsymbol{\mu}^{\ast}}(t) \simeq \varrho(t)\) (see Supplementary Materials). Last, transforming this state into momentum space provides \(\rho_{k}(t)\), from which the Uhlmann connection \(A_{\mathrm{U}}(k,t)\) is constructed. This allows a direct evaluation of the time-dependent Uhlmann geometric phase \(\Phi_{\mathrm{U}}(t)\) according to Eq.~\eqref{eq.Uhlmann}, providing a faithful measurement of the mixed-state topological invariant in open quantum systems.

The measured Uhlmann geometric phases are shown as green opaque bars in Fig.~\ref{fig.together}d, which agree well with the theoretical predictions (transparent bars). Near the critical time \(t_{c}\), where the driven Hamiltonian \(H_{\text{eff}}(t)\) changes its topology, the Uhlmann geometric phase exhibits a jump from 0 to \(\pi\). Immediately after \(t_{c}\), \(\Phi_{\mathrm{U}}(t)\) exhibits transient oscillations arising from residual coherence in the instantaneous eigenbasis, given the finite dephasing strength for each time step (see Supplementary Materials). At later times (\(t \gg t_{c}\)), the state becomes sufficiently dephased and \(\Phi_{\mathrm{U}}(t)\) stabilizes at the quantized value of \(\pi\). For the final state at \(t = 13\), we obtain \(\Phi_{\mathrm{U}}(t_{f})/\pi = 0.951 \pm 0.016\). The near-quantized value close to \(\pi\) is consistent with the nontrivial topological character of the final Hamiltonian, in sharp contrast to the topological mismatch between the Hamiltonian and the state observed in unitary dynamics (Fig.~\ref{fig.together}b). A quantitative comparison between the experimentally reconstructed states and the expected time-evolving mixed states during the noisy quench dynamics is provided in section C of the Supplementary Materials.

\subsection*{Observing topological kink structures}
The physical origin of the critical change in the quantized Uhlmann geometric phase \(\Phi_{\mathrm{U}}(t)\) is closely connected to the topological kink structures in the Uhlmann holonomy \cite{Viyuela2014,Viyuela2015}. To experimentally reveal this behavior, we extract the Uhlmann geometric phase \(\Phi_{\mathrm{U}}^{k} = \arg \lbrack\operatorname{Tr}(\rho_{k_{0}}\,\mathcal{P}\exp(\int_{k_{0}}^{k}A_{\mathrm{U}}))\rbrack\) for different \(k\), acquired during parallel transport across the first BZ from \(k_{0} = - \pi\) to \(k_{f} = \pi\). In Fig.~\ref{fig.kink}, we plot the measured \(\Phi_{\mathrm{U}}^{k}\) for three time-evolving mixed states under noisy quench dynamics. As shown in Fig.~\ref{fig.kink}a, for the initial state in the topologically trivial regime, \(\Phi_{\mathrm{U}}^{k}(t = 0) \simeq 0\) along the loop \(k \in \lbrack - \pi,\pi\rbrack\) and no topological kink structure arises. Immediately after the critical time \(t_{c}\), for a mixed state in the nontrivial regime, a critical momentum \(k_{c}\) emerges near the end points of the loop \(k_{f} = \pi\), as shown in Fig.~\ref{fig.kink}b. This critical momentum \(k_{c}\) divides the Uhlmann holonomy into two topologically inequivalent sectors, with \(\Phi_{\mathrm{U}}^{k} \simeq 0\) for \(k < k_{c}\) and \(\Phi_{\mathrm{U}}^{k} \simeq \pi\) for \(k > k_{c}\). At later times, the critical momentum \(k_{c}\) associated with the topological kink structures is observed to shift along \(k\) space, and the mixed states remain topologically nontrivial.

\subsection*{Probing the effect of quench velocity}
To further examine the robustness of the nontrivial topology of the dephased final state and clarify the role of the quench velocity, we implement an additional quench trajectory from the trivial initial Hamiltonian \((\theta_{1}^{i},\theta_{2}^{i}) = (2\pi/3,7\pi/8)\) to a distinct target Hamiltonian \((\theta_{1}^{f},\theta_{2}^{f}) = (\pi,0)\) within the nontrivial region according to \(\theta_{1,2}(t) = \theta_{1,2}^{i} + (\theta_{1,2}^{f} - \theta_{1,2}^{i})\upsilon t\), as presented in Fig.~\ref{fig.velocity}a. We experimentally vary the quench velocity \(v\) and reconstruct the corresponding dephased final states. As shown in Fig.~\ref{fig.velocity}b, the Uhlmann geometric phase of the reconstructed final states takes the nontrivial quantized value \(\Phi_{\mathrm{U}}(t_{f}) = \pi\) for sufficiently slow quench velocities (blue region), whereas it takes the trivial value of 0 as the quench velocity increases. This velocity-dependent behavior can be understood in light of Eq.~\eqref{eq.phase-limit}. For a fast quench with \(v = 1.0\), the excitation probability to the upper band is broadly distributed in momentum space (see Fig.~\ref{fig.velocity}c), indicating that the correction term \(\mathcal{O}(v^{1/2})\) becomes appreciable. Consequently, the dephased final state exhibits a trivial Uhlmann geometric phase. By contrast, for a slow quench with \(v = 0.1\), the excitations are confined to a narrower region around the energy-gap--closing point \(k = \pi\) (see Fig.~\ref{fig.velocity}d), while the lower band is predominantly occupied throughout most of the BZ. Then, the resulting Uhlmann geometric phase of dephased final state corresponds to the Berry phase of the final target Hamiltonian in the almost-adiabatic limit.

\section*{Discussion}

In summary, we have realized a direct measurement of the mixed-state topological invariant in Uhlmann's sense. Our approach is experimentally implemented in an open photonic QW by fully reconstructing the density matrix over the BZ. With this, we demonstrated that introducing dephasing noise into the slow-quench dynamics enables access to a topologically nontrivial state starting from an easily prepared trivial one. In addition, a topological kink structure in the Uhlmann holonomy arises after this topological phase transition. The nontrivial state behaves as a pseudoground state of the nontrivial Hamiltonian, with only a small density of excitations generated near the gap-closing points. The underlying mechanism---namely, the restoration of dynamically broken symmetries through noise---applies broadly to other symmetry classes of topological insulators and superconductors driven out of equilibrium \cite{McGinley2018,McGinley2019,Reid2022}. Moreover, our method for measuring mixed-state topological invariants, as well as for preparing topologically nontrivial mixed states via nonunitary dynamics, can be readily extended to open 2D topological systems. In this case, the relevant invariant is the topological Uhlmann number \cite{Viyuela2015,Viyuela2014a,Huang2014}, defined as the winding number of the Uhlmann geometric phase.

Quantum state tomography provides a well-established and information-rich route for characterizing topological properties and has been used across diverse physical platforms to reveal topologically inequivalent states, track the evolution of geometric phases and winding numbers, and determine complex topological charges \cite{Reid2022,Peng2014,Lin2026}. This strategy is complementary to approaches that infer a topological invariant from a suitably chosen observable or response through an established correspondence, whose applicability may depend on the Hamiltonian structure, protecting symmetries, initial state, or measurement protocol \cite{Cardano2017,Aidelsburger2015}. In our work, the reconstructed mixed states provide access not only to the quantized Uhlmann geometric phase but also to the energy-band populations, interband coherence, and the topological kink structure associated with the Uhlmann holonomy. Together, these quantities clarify the physical mechanism underlying the nonunitary preparation of a topologically nontrivial mixed state from an initially trivial state, as well as the associated transition of the quantized Uhlmann geometric phase.

While the tomographic method has been widely used to characterize pure-state topological phases, our work applies this strategy to high-dimensional open quantum systems and uses it to determine mixed-state topological properties. As the Hilbert-space dimension increases, traditional full-state tomography for topological characterizations can require substantial measurement and computational resources. In our experiment, the neural density operator ansatz provides an efficient variational representation of the high-dimensional mixed state and enables its high-fidelity reconstruction from experimentally measured data. This neural-network--assisted approach alleviates the resource demand associated with traditional tomographic reconstruction and makes the characterization of mixed-state topology in high-dimensional open systems more practical \cite{Carleo2019,Torlai2020}. With suitable state representations and experimentally accessible measurement data, this strategy can also be adapted to other open quantum simulators \cite{Altman2021}, providing a broadly applicable route for investigating mixed-state topological properties.

\section*{Materials and Methods}

\subsection*{Topological discrepancy between the time-dependent Hamiltonian and the state}
Here, we analytically demonstrate that a topological discrepancy naturally arises between the time-dependent Hamiltonian and the actual time-evolving state during a parameter quench. The effective Hamiltonian of the split-step QW at discrete time \(t \in \mathbb{N}\) takes the form \cite{Kitagawa2010,Asboth2013}

\begin{equation}\label{eq.Hamiltonian-time}
\begin{aligned}H(t) &= \int_{- \pi}^{\pi}{dk}H_{k}(t) \\ &= \int_{- \pi}^{\pi}{dk}\lbrack E_{k}(t)\mathbf{n}_{k}(t) \cdot \overrightarrow{\sigma}\rbrack \otimes |k\rangle\langle k|,\end{aligned}
\end{equation}

where discrete time \(t \in \mathbb{N}\), \(\overrightarrow{\sigma} = (\sigma_{x},\sigma_{y},\sigma_{z})\) is the Pauli vector. The energy \(E_{k}(t)\) and the Bloch vector \(\mathbf{n}_{k}(t)\) are parametrized by the two spin-rotation angles \(\{\theta_{1}(t),\theta_{2}(t)\}\), satisfying

\begin{equation}\label{eq.dispersion}
\begin{aligned}
\cos \lbrack E_{k}(t)\rbrack &= \cos \lbrack\frac{\theta_{1}(t)}{2}\rbrack\cos \lbrack\frac{\theta_{2}(t)}{2}\rbrack\cos k \\ &\quad - \sin \lbrack\frac{\theta_{1}(t)}{2}\rbrack\sin \lbrack\frac{\theta_{2}(t)}{2}\rbrack,
\end{aligned}
\end{equation}

and
\begin{equation}\label{eq.Bloch-vector}
\left\{ \begin{aligned}
\mathbf{n}_{k}^{x}(t) & =0, \\
\mathbf{n}_{k}^{y}(t) & =\frac{\begin{gathered}\sin \left\lbrack \frac{\theta_{1}(t)}{2} \right\rbrack\cos \left\lbrack \frac{\theta_{2}(t)}{2} \right\rbrack\cos k \\ + \cos \left\lbrack \frac{\theta_{1}(t)}{2} \right\rbrack\sin \left\lbrack \frac{\theta_{2}(t)}{2} \right\rbrack\end{gathered}}{\sin \left\lbrack E_{k}(t) \right\rbrack}, \\
\mathbf{n}_{k}^{z}(t) & = - \frac{\cos \left\lbrack \frac{\theta_{2}(t)}{2} \right\rbrack\sin k}{\sin \left\lbrack E_{k}(t) \right\rbrack},
\end{aligned} \right.
\end{equation}

respectively.

As an experimentally relevant setting, we consider an initial state prepared as the ground state of a trivial Hamiltonian \(H(\theta_{1}^{i},\theta_{2}^{i})\), i.e., within the regime \(|\text{tan}(\theta_{2}^{i}/2)/\text{tan}(\theta_{1}^{i}/2)| > 1\). The control parameters are then slowly quenched across the topological phase boundary toward a target nontrivial Hamiltonian \(H(\theta_{1}^{f},\theta_{2}^{f})\) satisfying \(|\text{tan}(\theta_{2}^{f}/2)/\text{tan}(\theta_{1}^{f}/2)| < 1\). We use the Berry phase (or Zak phase in one dimension) to characterize the topological properties of both the time-dependent Hamiltonian \(H(\theta_{1}(t),\theta_{2}(t))\) and the time-evolving state \(\Phi_{\mathrm{B}}(t)\). For ground state of \(H(\theta_{1}(t),\theta_{2}(t))\), the Berry phase \(\Phi_{\mathrm{B}}^{H}\) changes from 0 to \(\pi\) when the Hamiltonian crosses the phase boundary. By contrast, the Berry phase of the initial state is \(\Phi_{\mathrm{B}}(t = 0) = i\int_{- \pi}^{\pi}\langle\psi_{k}(0)|\partial_{k}|\psi_{k}(0)\rangle = 0\), where \(|\psi_{k}(0)\rangle = |\mathbf{n}_{k}(0)\rangle\) is the eigenstate of the initial Bloch Hamiltonian \(H_{k}(\theta_{1}^{i},\theta_{2}^{i})\) with eigenvalue of -1. After a one-step unitary evolution, the Berry phase of the instantaneous state becomes

\begin{equation}\label{eq.Berry-unitary}
\begin{aligned}
\Phi_{\mathrm{B}}(t = 1)&=  i\int_{- \pi}^{\pi}{dk}\langle\psi_{k}(0)|U_{k}^{\dagger}\partial_{k}(U_{k}|\psi_{k}(0)\rangle) \\
&=  \Phi_{\mathrm{B}}(0) + i\int_{- \pi}^{\pi}{dk}\operatorname{Tr}\{ \\ &\quad \frac{1}{2}\lbrack\mathbb{1} - \mathbf{n}_{k}(0) \cdot \overrightarrow{\sigma}\rbrack \\ &\quad \lbrack\mathbb{1}\cos(E_{k}) + i\sin (E_{k})\mathbf{n}_{k} \cdot \overrightarrow{\sigma}\rbrack \\
&\quad \times \lbrack - \mathbb{1}\sin(E_{k})\partial_{k}(E_{k}) \\ &\qquad + i\cos (E_{k})\partial_{k}(E_{k})\mathbf{n}_{k} \cdot \overrightarrow{\sigma} \\ &\qquad + i\sin (E_{k})\partial_{k}(\mathbf{n}_{k}) \cdot \overrightarrow{\sigma}\rbrack\} \\
&=  0,
\end{aligned}
\end{equation}

where \(U_{k} = \mathrm{e}^{- iH_{k}}\), and \(\mathbb{1}\) is the identity matrix. The final equality follows from the fact that the integrand is an odd function over the symmetric interval \(\lbrack - \pi,\pi\rbrack\) and therefore integrates to zero. Hence, the Berry phase of the state is unchanged after the first unitary step. By induction, the PHS-protected Berry phase of the evolving state remains constant throughout the entire coherent quench, although the Hamiltonian itself undergoes a topological transition. This topological discrepancy prohibits the preparation of a topologically nontrivial state from an initially trivial one under unitary dynamics.

\subsection*{Dynamic-induced symmetry breaking and noise-assisted symmetry recovery}
We present an analytic argument showing that the antiunitary symmetries of a time-evolving state, although preserved by the instantaneous Hamiltonian, are generically violated during coherent quench dynamics and can, nonetheless, be restored by introducing dephasing noise. The essential reason for the mismatch between the geometric phase of the Hamiltonian, \(\Phi_{\mathrm{B}}^{H}(t)\), and that of the evolving state, \(\Phi_{\mathrm{B}}(t)\), in 1D QW is the phenomenon known as dynamically induced symmetry breaking \cite{McGinley2018,McGinley2019,Reid2022}. For QW in symmetry class BDI, CS, TRS, and PHS are all respected by the Hamiltonian throughout the coherent quench. However, the state itself fails to obey TRS and CS under unitary evolution. For the momentum-resolved density matrix \(\rho_{k}(t)\), one finds the antiunitary operators \(\mathrm{T}\) and \(\Gamma\) satisfying

\begin{equation}\label{eq.TRS-breaking}
\begin{aligned}
\mathrm{T}\rho_{k}(t)\mathrm{T}^{\dagger}&=  \sigma_{z}\rho_{k}(t)\sigma_{z} \\
&=  \rho_{k}^{\ast}( - t) \neq \rho_{k}^{\ast}(t),
\end{aligned}
\end{equation}

and
\begin{equation}\label{eq.CS-breaking}
\begin{aligned}
\Gamma\rho_{k}(t)\Gamma^{\dagger}&=  ( - i\sigma_{x})\rho_{k}(t)(i\sigma_{x}) \\
&=  1 - \rho_{k}( - t) \neq 1 - \rho_{k}(t),
\end{aligned}
\end{equation}

respectively, demonstrating the breakdown of TRS and CS in the time-evolving state.

To counteract this dynamically induced symmetry breaking, we introduce classical fluctuations that act as pure dephasing noise in the instantaneous eigenbasis of the Bloch Hamiltonian \(H_{k}(t)\). The corresponding evolution equation is given by \cite{Ying2016}

\begin{equation}\label{eq.master}
\dot{\rho}_{k} = - i\lbrack H_{k},\rho_{k}\rbrack + \gamma_{k}\lbrack{\widetilde{\sigma}}_{k}^{z}\rho_{k}{\widetilde{\sigma}}_{k}^{z} - \rho_{k}\rbrack,
\end{equation}

where ``\(\cdot\)'' means the derivative of time \(t\), \(\gamma_{k}\) is the momentum-resolved dephasing rate, and \({\widetilde{\sigma}}_{k}^{z}(t) = \mathbf{n}_{k}(t) \cdot \overrightarrow{\sigma}\) is the Pauli operator in the eigenbasis of \(H_{k}(t)\). In the strong-dephasing limit \(\gamma_{k}t_{f} \gg 0\), the final state at momentum \(k\) becomes fully diagonal in the eigenbasis of the final Hamiltonian

\begin{equation}\label{eq.dephased-populations}
\rho_{k}^{\text{de}} = \frac{1}{2}\lbrack\mathbb{1} + (2P_{g \rightarrow e}^{k} - 1)\mathbf{n}_{k}(t_{f}) \cdot \overrightarrow{\sigma}\rbrack,
\end{equation}

with \(P_{g \rightarrow e}^{k}\) the excitation probability from the lower energy band to the upper one accumulated as the quench passes the gap-closing time \(t_{c}\). The Landau-Zener expression provides a good approximation for the excitation probability: \(P_{g \rightarrow e}^{k} \approx \mathrm{e}^{- \frac{2\pi{|\epsilon_{12}|}^{2}}{\beta}}\). Here, \(\epsilon_{12}\) is the off-diagonal term of the Hamiltonian, and \(\beta = |d(\epsilon_{2} - \epsilon_{1})/dt|\) is the energy gap changing rate near the phase-transition time \(t_{c}\). In the limit of a slow quench, the excitation only happens near the gap-closing point \(k = 0\), so we have \(\beta = \left| 2\dot{E}_{k}(t) \right| \approx \left| \dot{\theta}_{1}(t) + \dot{\theta}_{2}(t) + \mathcal{O}\left( k^{2} \right) \right|\), where \(\mathcal{O}\) denotes the order of the function. Moreover, near the critical time \(t_{c}\), \(\dot{\theta}_{1(2)}(t) = \dot{\theta}_{1(2)}(t_{c}) + \mathcal{O}(t - t_{c})\), and we lastly get \(\beta = |\upsilon + \mathcal{O}(k^{2}) + \mathcal{O}(t - t_{c})|\), where \(\upsilon = \dot{\theta}_{1}(t_{c}) + \dot{\theta}_{2}(t_{c})\). By a similar approximation process, we have

\begin{equation}\label{eq.coupling}
\begin{aligned}
{|\epsilon_{12}|}^{2}&=  |\langle\mathbf{n}_{k}^{\text{u}}(t_{f})|E_{k}(t)\mathbf{n}_{k}(t) \cdot \overrightarrow{\sigma}{|\mathbf{n}_{k}^{\mathrm{d}}(t_{f})\rangle|}^{2} \\
&\approx  \frac{k^{2}(1 + \cos \theta_{1}(t_{c}))}{2} + \mathcal{O}(k^{3}) \\ &\quad + \mathcal{O}(t - t_{c}),
\end{aligned}
\end{equation}

where \(|\mathbf{n}_{k}^{\text{u(d)}}(t_{f})\rangle\) is the eigenstate of the upper (lower) energy band of the final Bloch Hamiltonian \(H_{k}(t_{f})\). As a result, we get that

\begin{equation}\label{eq.LZ}
P_{g \rightarrow e}^{k} = \mathrm{e}^{- \frac{\pi k^{2}(1 + \cos \theta_{1}(t_{c}))}{|\upsilon|}}.
\end{equation}

Because the dephased state \(\rho_{k}^{\text{de}}\) is diagonal in the eigenbasis of the final Hamiltonian \(H_{f}\), it thus would have the same symmetry as the Hamiltonian \(H_{f}\). Thus, the addition of dephasing noise can restore the symmetries that were dynamically broken during the unitary evolution, resolving the symmetry mismatch responsible for the topological discrepancy.

\subsection*{Dynamical preparation of a topologically nontrivial mixed state}
We show that the geometric phase of the dephased mixed state in Eq.~\eqref{eq.dephased-populations} matches the topological character of the final Hamiltonian. To characterize the topology of mixed states resulting from noisy quench dynamics, we adopt the definition of the Uhlmann geometric phase \cite{Viyuela2014}. Following the theory of Uhlmann \cite{Uhlmann1986}, any mixed state \(\rho_{\mathbf{k}}\) admits a purification of the form \(\rho_{\mathbf{k}} = \omega_{\mathbf{k}}\omega_{\mathbf{k}}^{\dagger}\), where \(\omega_{\mathbf{k}} = \sqrt{\rho_{\mathbf{k}}}U_{\mathbf{k}}\text{ and }U_{\mathbf{k}}\) is arbitrary. This freedom defines a \(U(n)\) gauge structure for an \(n\)-dimensional density matrix. Analogous to the \(U(1)\) parallel-transport condition for pure states, the Uhlmann parallel-transport condition is: \({\overset{.}{\omega}}_{\mathbf{k}}^{\dagger}\omega_{\mathbf{k}} = \omega_{\mathbf{k}}^{\dagger}{\overset{.}{\omega}}_{\mathbf{k}}\). If \(\rho_{\mathbf{k}} = \sum_{i}p_{\mathbf{k}}^{i}|\phi_{\mathbf{k}}^{i}\rangle\langle\phi_{\mathbf{k}}^{i}|\) is full rank (i.e., \(p_{\mathbf{k}}^{i} \neq 0\)), the cyclic parallel transport operator is \(V = \mathrm{e}^{\oint A_{\mathrm{U}}}\), where \(A_{\mathrm{U}} = \sum_{i,j}\frac{|\phi_{\mathbf{k}}^{i}\rangle\langle\phi_{\mathbf{k}}^{i}|\lbrack\partial_{\mathbf{k}}\sqrt{\rho_{\mathbf{k}}},\sqrt{\rho_{\mathbf{k}}}\rbrack|\phi_{\mathbf{k}}^{j}\rangle\langle\phi_{\mathbf{k}}^{j}|}{p_{\mathbf{k}}^{i} + p_{\mathbf{k}}^{j}}dk\). Under a closed parameter loop, this evolution maps \(\omega_{\mathbf{k}(0)}\overset{V}{\rightarrow}\omega_{\mathbf{k}(1)}\), where \(\omega_{\mathbf{k}(0)}\) and \(\omega_{\mathbf{k}(1)}\) are the start and end points of the parallel transport, respectively. The geometric phase defined by this Uhlmann parallel transport thus can be calculated as

\begin{equation}\label{eq.Uhlmann-general}
\begin{aligned}
\Phi_{\mathrm{U}} &= \arg (\operatorname{Tr}\lbrack\omega_{\mathbf{k}(0)}^{\dagger}\omega_{\mathbf{k}(1)}\rbrack) \\ &= \arg (\operatorname{Tr}\lbrack\rho_{\mathbf{k}(0)}\mathcal{P}\mathrm{e}^{\oint A_{\mathrm{U}}}\rbrack),
\end{aligned}
\end{equation}

where \(\rho_{\mathbf{k}(0)}\) is the mixed state at the start point of the cyclic parallel transport, \(\arg (x)\) means the function to get the argument of \(x\), and \(\mathcal{P}\) denotes path ordering.

For the 1D QW considered here, \(\mathbf{k} \equiv k\) is a 1D lattice momentum. In its spectral representation, \(\rho_{k} = \sum_{i = \pm}^{}p_{k}^{i}|\phi_{k}^{i}\rangle\langle\phi_{k}^{i}|\), the Uhlmann connection \(A_{\mathrm{U}}\) becomes

\begin{equation}\label{eq.connection}
\begin{aligned}
A_{\mathrm{U}}&=  \frac{1}{2}(\mathbb{1} + \mathbf{n}_{k} \cdot \overrightarrow{\sigma})\lbrack\frac{iN_{k}^{2}}{2}(\partial_{k}\mathbf{n}_{k} \times \mathbf{n}_{k}) \cdot \overrightarrow{\sigma}\rbrack \\ &\quad \frac{1}{2}(\mathbb{1} - \mathbf{n}_{k} \cdot \overrightarrow{\sigma})dk \\
&\quad + \frac{1}{2}(\mathbb{1} - \mathbf{n}_{k} \cdot \overrightarrow{\sigma})\lbrack\frac{iN_{k}^{2}}{2}(\partial_{k}\mathbf{n}_{k} \times \mathbf{n}_{k}) \cdot \overrightarrow{\sigma}\rbrack \\ &\quad \frac{1}{2}(\mathbb{1} + \mathbf{n}_{k} \cdot \overrightarrow{\sigma})dk \\
&=  \frac{iN_{k}^{2}}{2}(\partial_{k}\mathbf{n}_{k} \times \mathbf{n}_{k}) \cdot \overrightarrow{\sigma}dk,
\end{aligned}
\end{equation}

where \(N_{k} = \frac{\sqrt{1 + r_{k}} - \sqrt{1 - r_{k}}}{\sqrt{2}}\) with \(r_{k} = p_{k}^{+} - p_{k}^{-}\), and \(\mathbf{n}_{k}\) is the normalized Bloch vector of the mixed state. Hence, the Uhlmann phase of the mixed state reads

\begin{equation}\label{eq.phase-reduction}
\begin{aligned}
\Phi_{\mathrm{U}}&=  \arg (\operatorname{Tr}\lbrack\rho_{- \pi}\mathrm{e}^{\int_{- \pi}^{\pi}dk\,\frac{iN_{k}^{2}}{2}(\partial_{k}\mathbf{n}_{k} \times \mathbf{n}_{k}) \cdot \overrightarrow{\sigma}}\rbrack) \\
&=  \arg (\operatorname{Tr}\lbrack\frac{1}{2}(\mathbb{1} - \sigma_{y}) \\ &\quad (\mathbb{1}\cos(|\overrightarrow{U}|) + i \ast \mathbf{n}_{\mathrm{U}} \cdot \overrightarrow{\sigma}\sin (|\overrightarrow{U}|))\rbrack) \\
&=  \arg (\cos \lbrack|\overrightarrow{U}|\rbrack),
\end{aligned}
\end{equation}

where \(\overrightarrow{U} = \int_{- \pi}^{\pi}{dk}\frac{N_{k}^{2}}{2}(\partial_{k}\mathbf{n}_{k} \times \mathbf{n}_{k})\) and \(\mathbf{n}_{\mathrm{U}} = \overrightarrow{U}/|\overrightarrow{U}|\).

For the fully dephased state \(\rho_{k}^{\text{de}}\) in Eq.~\eqref{eq.dephased-populations}, the Bloch-vector length is \(r_{k} = |2\mathrm{e}^{- \frac{\alpha k^{2}}{|\upsilon|}} - 1|\) with \(\alpha = \pi\lbrack 1 + \cos \theta_{1}(t_{c})\rbrack\). It is clear that when \(\frac{\alpha k^{2}}{|\upsilon|} \gg 1\), \(r_{k} = 1\). The modulus of \(\overrightarrow{U}\) thus becomes

\begin{equation}\label{eq.holonomy-split}
\begin{aligned}
|\overrightarrow{U}|&=  \int_{\frac{\alpha k^{2}}{|\upsilon|} \gg 1}^{}{dk}\frac{1}{2}{(\partial_{k}\mathbf{n}_{k} \times \mathbf{n}_{k})}_{x} \\ &\quad + \int_{\text{other}}^{}{dk}\frac{N_{k}^{2}}{2}{(\partial_{k}\mathbf{n}_{k} \times \mathbf{n}_{k})}_{x} \\
&=  |\int_{- \pi}^{\pi}{dk}\frac{1}{2}{(\partial_{k}\mathbf{n}_{H}^{k} \times \mathbf{n}_{H}^{k})}_{x} \\ &\quad + \int_{\text{other}}^{}{dk}\frac{N_{k}^{2} - 1}{2}{(\partial_{k}\mathbf{n}_{H}^{k} \times \mathbf{n}_{H}^{k})}_{x}| \\
&=  |\Phi_{\mathrm{B}}^{H_{f}} \\ &\quad + \int_{\text{other}}^{}{dk}\frac{N_{k}^{2} - 1}{2}{(\partial_{k}\mathbf{n}_{H}^{k} \times \mathbf{n}_{H}^{k})}_{x}|,
\end{aligned}
\end{equation}

where \(\frac{1}{2}\int_{- \pi}^{\pi}dk{(\partial_{k}\mathbf{n}_{H}^{k} \times \mathbf{n}_{H}^{k})}_{x} = \Phi_{\mathrm{B}}^{H_{f}}\) is the Berry geometric phase of the final Hamiltonian \(H_{f}\) with chiral operator \(\Gamma = \sigma_{x}\). Moreover, for the completely dephased state, the second term is approximate as

\begin{equation}\label{eq.slow-quench-correction}
\begin{aligned}
\text{ } & \int_{- \sqrt{\frac{\beta\upsilon}{\alpha}}}^{\sqrt{\frac{\beta\upsilon}{\alpha}}}{dk}\,\frac{N_{k}^{2} - 1}{2}{(\partial_{k}\mathbf{n}_{k} \times \mathbf{n}_{k})}_{x} \\
&\approx  \int_{- \sqrt{\frac{\beta\upsilon}{\alpha}}}^{\sqrt{\frac{\beta\upsilon}{\alpha}}}{dk}\,\lbrack - \sqrt{\frac{\alpha}{|\upsilon|}}|k| + \mathcal{O}({|k|}^{3})\rbrack \\ &\quad \lbrack\cos \frac{\theta_{2}}{2}\csc \frac{\theta_{1} + \theta_{2}}{2}\mathcal{O}(k^{2})\rbrack \\
&\approx  \mathcal{O}(\upsilon^{\frac{1}{2}}),
\end{aligned}
\end{equation}

where \(\mathrm{e}^{- \beta} \ll 1\). From the above, we have \(|\overrightarrow{U}| \approx |\Phi_{\mathrm{B}}^{H_{f}} + \mathcal{O}(\upsilon^{\frac{1}{2}})|\). At least for small \(\upsilon\), we have \(|\overrightarrow{U}| > \pi/2\) (or \(|\overrightarrow{U}| < \pi/2\)), leading to a quantized Uhlmann geometric phase: \(\Phi_{\mathrm{U}}^{\text{de}} = \pi\) (or 0). Hence, a topologically nontrivial mixed state with \(\Phi_{\mathrm{U}}^{\text{de}} = \pi\) can be obtained through a noisy quench from a trivial Hamiltonian into the target nontrivial Hamiltonian.

\subsection*{Effective dephasing mechanism of ensemble-averaged dynamics}
We show how the fluctuation amplitude of the rotation parameter gives rise to an effective dephasing rate in the ensemble-averaged dynamics. In the instantaneous eigenbasis of the time-dependent Hamiltonian \(H_{k}(t)\), the Bloch Hamiltonian is diagonal and reads as

\begin{equation}\label{eq.spectral-Hamiltonian}
H_{k}(t) = E_{k}(t)\mathbf{n}_{k}(t) \cdot \overrightarrow{\sigma} = E_{k}(t){\widetilde{\sigma}}_{k}^{z}(t).
\end{equation}

We denote the instantaneous eigenstates of \(H_{k}(t)\) by \(\{|e_{k}(t)\rangle,|g_{k}(t)\rangle\}\), corresponding to the eigenvalues \(\pm E_{k}(t)\), respectively. In this instantaneous eigenbasis, the matrix elements of the time-evolving density matrix can be written as \(\rho_{\alpha\beta}^{k}(t) = \langle\alpha_{k}(t)|\rho_{k}(t)|\beta_{k}(t)\rangle\) (\(\alpha,\beta = e,g\)). Here, \(\rho_{ee}^{k}\) and \(\rho_{gg}^{k}\) are the excited- and ground-state populations, whereas the off-diagonal elements \(\rho_{eg}^{k}\) and \(\rho_{ge}^{k}\) characterize the coherence between the two instantaneous eigenstates. In the instantaneous eigenbasis, \({\widetilde{\sigma}}_{k}^{z}(t)\) is diagonal. Therefore, the noisy term in Eq.~\eqref{eq.dephasing} acts on the time-evolving density matrix as

\begin{equation}\label{eq.noise-matrix}
\begin{aligned}
&\gamma_{k}(t)\left\lbrack {\widetilde{\sigma}}_{k}^{z}(t)\rho_{k}(t){\widetilde{\sigma}}_{k}^{z}(t) - \rho_{k}(t) \right\rbrack \\ &= \begin{pmatrix}
0 & - 2\gamma_{k}(t)\rho_{eg}^{k}(t) \\
 - 2\gamma_{k}(t)\rho_{ge}^{k}(t) & 0
\end{pmatrix}.
\end{aligned}
\end{equation}

This indicates that the noisy term does not change the diagonal elements of the density matrix \(\rho_{ee}^{k}\) and \(\rho_{gg}^{k}\) but only suppresses the off-diagonal elements. Thus, the noisy term with \({\widetilde{\sigma}}_{k}^{z}\) as the dephasing operator realizes pure dephasing between the ground and excited states. Combining the Hamiltonian evolution with the noisy term, the off-diagonal element \({\dot{\rho}}_{eg}^{k}(t) = - i\, 2E_{k}(t)\rho_{eg}^{k}(t) - 2\gamma_{k}(t)\rho_{eg}^{k}(t)\). The Hamiltonian term only contributes a dynamical phase factor, whereas the noisy term leads to the decay of the coherence. The solution for this off-diagonal element is

\begin{equation}\label{eq.coherence}
\rho_{eg}^{k}(t) = e^{- i\int_{0}^{t}2E_{k}(s)\mathrm{d}s}e^{- 2\int_{0}^{t}\gamma_{k}(s)\mathrm{d}s}\rho_{eg}^{k}(0).
\end{equation}

Because \(\rho_{k}(t)\) is Hermitian, the two off-diagonal elements satisfy \(\rho_{ge}^{k}(t) = {\lbrack\rho_{eg}^{k}(t)\rbrack}^{\ast}\).

In the experiment, the dephasing noise mechanism is effectively emulated by the ensemble-averaged dynamics under a slowly fluctuating rotation angle \cite{Schreiber2011,Tang2022}. For each disordered realization, the rotation parameter is randomly shifted as \(\theta_{1}(t) \rightarrow \theta_{1}(t) + \xi\), where \(\xi\) is uniformly sampled from the interval \(\lbrack - \delta\theta_{1},\delta\theta_{1}\rbrack\). Here, \(\delta\theta_{1}\) is the experimentally controlled fluctuation amplitude. Each disordered realization corresponds to a unitary QW evolution, and the effective nonunitary dynamics arises after ensemble averaging over many disordered realizations. To identify the effective dephasing, we expand the disordered Hamiltonian to first order as \(H_{k}(t;\xi) \equiv H_{k}\lbrack\theta_{1}(t) + \xi\rbrack \simeq H_{k}(t) + \xi\,\partial_{\theta_{1}}H_{k}(t)\). The perturbation term \(\partial_{\theta_{1}}H_{k} = (\partial_{\theta_{1}}E_{k})\mathbf{n}_{k} \cdot \overrightarrow{\sigma} + E_{k}(\partial_{\theta_{1}}\mathbf{n}_{k}) \cdot \overrightarrow{\sigma}.\) The first contribution of the perturbation term is diagonal in the instantaneous eigenbasis and changes only the eigenvalue difference between the two instantaneous eigenstates. The second contribution changes the instantaneous eigenbasis and may induce transitions between the two instantaneous eigenstates. For the parameter regime used in our experiment, the first contribution gives the dominant effect. This spectral-disorder--dominated approximation is numerically shown by fig. S1, where the ensemble-averaged dynamics agrees well with the ideal pure-dephasing dynamics in both the energy band occupation and the purity of the dephased density matrix in the first BZ.

Under the spectral-disorder--dominated approximation \cite{Kropf2016}, the eigenvalue difference between the two instantaneous eigenstates can be written as

\begin{equation}\label{eq.spectral-disorder}
\begin{aligned}
2E_{k}(t;\xi) &\simeq 2E_{k}(t) + 2\xi\partial_{\theta_{1}}E_{k}(t) \\ &= 2E_{k}(t) + \xi\partial_{\theta_{1}}\lbrack 2E_{k}(t)\rbrack.
\end{aligned}
\end{equation}

For a given disordered realization \(\xi\), this fluctuation of the eigenvalue difference changes the phase factor multiplying the off-diagonal elements \(\rho_{eg}^{k,\xi}(t)\), which evolves as

\begin{equation}\label{eq.random-coherence}
\rho_{eg}^{k,\xi}(t) = e^{- i\int_{0}^{t}2E_{k}(s)\mathrm{d}s}e^{- i\xi A_{k}(t)}\rho_{eg}^{k}(0),
\end{equation}

where \(A_{k}(t) = \int_{0}^{t}{\mathrm{d}s}\partial_{\theta_{1}}\lbrack 2E_{k}(s)\rbrack\) quantifies how the fluctuation of \(\theta_{1}\) is accumulated into a random phase \(\xi A_{k}(t)\). Thus, apart from the dynamical phase factor, the disordered realization contributes a random phase factor \(\exp \lbrack - i\xi A_{k}(t)\rbrack\). Averaging this random phase factor over the uniform distribution of \(\xi \in \lbrack - \delta\theta_{1},\delta\theta_{1}\rbrack\) gives the characteristic function

\begin{equation}\label{eq.sinc}
\begin{aligned}
\left\langle e^{- i\xi A_{k}(t)} \right\rangle_{\xi} &= \frac{1}{2\delta\theta_{1}}\int_{- \delta\theta_{1}}^{\delta\theta_{1}}\mathrm{d}\xi\, e^{- i\xi A_{k}(t)} \\ &=\operatorname{sinc}\left\lbrack \delta\theta_{1}A_{k}(t) \right\rbrack,
\end{aligned}
\end{equation}

where \(\operatorname{sinc}(x) = \sin x/x\). Compared to the purely dephasing dynamics given by Eq.~\eqref{eq.coherence}, the effective dephasing rate associated with the ensemble-averaged fluctuating dynamics is obtained from \(\exp \left\lbrack - 2\int_{0}^{t}\gamma_{k}(s)\mathrm{d}s \right\rbrack = \left| \operatorname{sinc}\left\lbrack \delta\theta_{1}A_{k}(t) \right\rbrack \right|\). Here, the modulus is used because the effective dephasing rate is determined by the decay of the magnitude of the ensemble-averaged off-diagonal element, rather than by its phase. Taking the logarithmic derivative with respect to time, we obtain the explicit relation

\begin{equation}\label{eq.effective-rate}
\gamma_{k}(t) = - \frac{1}{2}\frac{\mathrm{d}}{\mathrm{d}t}\ln \left| \operatorname{sinc}\left\lbrack \delta\theta_{1}A_{k}(t) \right\rbrack \right|.
\end{equation}

Thus, \(\delta\theta_{1}\) determines the width of the spectral-disorder distribution, while \(\gamma_{k}(t)\) is the effective time-dependent dephasing rate inferred from the ensemble-averaged dynamics. In the weak-fluctuation regime, \(|\delta\theta_{1}A_{k}(t)| \ll 1\), the effective dephasing rate reduces to

\begin{equation}\label{eq.weak-rate}
\gamma_{k}(t) \simeq \frac{\delta\theta_{1}^{2}}{6}A_{k}(t)\partial_{\theta_{1}}\lbrack 2E_{k}(t)\rbrack.
\end{equation}

In this weak-fluctuation regime, the effective dephasing rate scales quadratically with the fluctuation amplitude of the rotation parameter.

\bibliographystyle{apsrev4-1PRX}
\renewcommand{\bibsection}{\section*{References}}
\bibliography{referenceV1}

\end{document}